\documentclass[aps,nofootinbib,twocolumn,floatfix]{revtex4}
\usepackage{graphicx}
\graphicspath{{figures/}} 
\usepackage{amsmath}
\usepackage{amsfonts}
\usepackage{amssymb}
\usepackage{dcolumn}
\usepackage{braket}
\usepackage{commath}
\usepackage{epstopdf}
\usepackage{cancel}
\usepackage{tabularx, booktabs}
\usepackage[labelsep=quad,indention=0pt]{subfig}
\usepackage{tikz}
\usetikzlibrary{calc,positioning,fit,backgrounds}
\usepackage{pgfplots}

\usepackage{color}

\begin{document}

\title{The tight-binding formulation of the Kronig-Penney model}
\author{F. Marsiglio and R.~L.~Pavelich}
\email{fm3@ualberta.ca, rpavelic@ualberta.ca}
\affiliation{Department of Physics, University of Alberta, Edmonton, AB, Canada T6G 2E1}

\begin{abstract}
We provide a derivation of the tight-binding model that emerges from a full consideration of
a particle bound in a periodic one-dimensional array of square well potentials, separated by barriers
of height $V_0$ and width $b$. We derive the dispersion for such a model, and show that an effective
next-nearest-neighbor hopping parameter is required for an accurate description. An electron-hole
asymmetry is prevalent except in the extreme tight-binding limit, and emerges through a ``next-nearest neighbor''
hopping term in the dispersion. We argue that this does not necessarily imply next-nearest-neighbor tunneling; this is
done by deriving the transition
amplitudes for a two-state effective model that describes a double-well potential, which is a simplified precursor
to the problem of a periodic array of potential wells.
\end{abstract}

\date{\today} 
\maketitle

\section{INTRODUCTION}
\label{sec:intro}
The notion of an ``effective model" or ``effective potential" pervades essentially all of physics. 
At the undergraduate level, for example, it is worth
emphasizing that even the lowly harmonic oscillator potential is really merely an ``effective potential." In reality all potentials
are generally more complicated---the spring will eventually stretch inelastically---and any potential has a useful
domain of applicability in every problem. We have already emphasized this approach to simple one-body potentials
through the use of containment within a ``universe," i.e.~the one-dimensional harmonic oscillator potential contained within
an infinite square well, where, for low-lying states, this (extreme) deviation from harmonicity was shown to {\it not}  effect
the eigenvalues or eigenstates.\cite{marsiglio09} This was further illustrated with double-well potentials,\cite{jelic12,dauphinee15}
and with periodic potentials.\cite{pavelich15,pavelich16}

One can go further with ``effective models," with perhaps the best-known example being Feynman's
description of the ammonia molecule as a two-state system.\cite{feynman65}
This sort of description is worthwhile for certain aspects
of the problem, such as the time dependence of the wave function, but it remains unclear how parameters required in the effective model
are related to underlying ``microscopic" characteristics of the same problem. A more concrete example is the double-well potential. In Ref.~[\onlinecite{dauphinee15}] the states describing a particle in such a system were determined from the basic
parameters, namely the barrier height and width. At the same time, a ``toy model'' involving a single parameter, a transition amplitude $t$,
to describe
tunneling through the barrier, was shown to very accurately describe the ground state
splitting calculated by solving the complete Schr\"odinger
equation. This was a case in point, where the original model had a Hilbert space of infinite dimension, the ``toy'' or ``effective'' model
was only a two-state system. While Ref.~[\onlinecite{dauphinee15}] provided an estimate for the transition amplitude $t$ in terms of the
microscopic parameters of the original model, the correspondence was only approximate.

Another example that occurs in condensed matter is that of band structure calculations, where, in the simplest case, a periodic
array of some potential gives rise to energy bands, whose characteristics require a microscopic solution to the Schr\"odinger
equation. This calculation in principle involves an infinite Hilbert space, in two senses. First, even a single potential well, representing
a single atom with which an electron interacts, requires an infinite Hilbert space. However, for a solid there are a large number of these
wells---an infinite number if we allow the solid to go on forever. This latter infinity is handled analytically through 
Bloch's theorem,\cite{bloch29, ashcroft76, pavelich15} which allows solution of the electron wave function in the infinite periodic 
array in terms of the solution within a single well (or unit cell, to use the jargon of condensed matter). A detailed modern description
of Bloch's theorem is given in Ref.~[\onlinecite{ashcroft76}] and a simplified description in terms of the currently defined problem is given
in Ref.~[\onlinecite{pavelich15}]. Even with Bloch's theorem, however, an infinite Hilbert space is required to describe the (infinite) set
of energy bands that emerge from the periodicity. An effective model, known as a ``tight-binding model,''
reduces the infinite Hilbert space down to an $N$-dimensional Hilbert space, where $N$ is the number of atoms. This description
is entirely analogous to the reduction of the double-well potential to a 2-dimensional Hilbert space, involving a single tunneling parameter
$t$, and in fact, even as $N \rightarrow \infty$, only a single parameter $t$, called the ``tunneling amplitude,'' remains.

The purpose of this paper is to use the one-dimensional Kronig-Penney model\cite{kronig31} to derive expressions for the 
one or two parameters in the ``effective" tight-binding model, in terms of the parameters that describe the original Kronig-Penney
model. This calculation is possible because some aspects of the one-dimensional Kronig-Penney model are known analytically,
and will complement the phenomenological fits realized in Refs.~[\onlinecite{pavelich16,dauphinee15}]. The
latter reference describes a double-well potential, and can be thought of as a special preliminary case of the fully periodic solid.
We begin with a brief description of this case,
and then describe the derivation in full
in the case of the Kronig-Penney model. Another double-well solution, more pertinent to the Kronig-Penney model, is briefly
described in Appendix A. For the Kronig-Penney model, our numerical solutions confirm that this description is exact in the limit
of tightly-bound wells, and we explore further to what degree this description remains accurate as the coupling between wells
increases. This exercise, with a specific model,\cite{kronig31} should result in a deeper understanding of 
the connection between effective models and their more microscopic counterparts.

\section{Double-well Potential}
\label{sec:doubwell}

We begin the the double-well potential, discussed generically in Ref.~[\onlinecite{jelic12}] and more specifically in 
Ref.~[\onlinecite{dauphinee15}]. 
\begin{figure}[h]
\includegraphics[width=5.0cm,angle=-90]{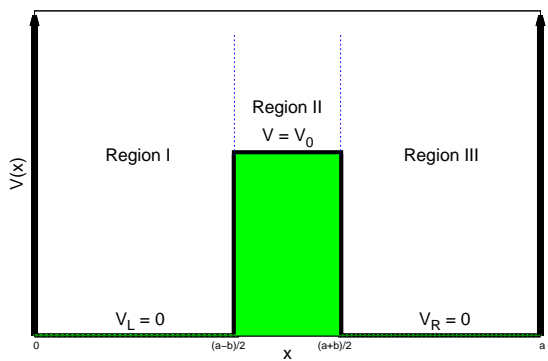}
\caption{Schematic of double-well potential.
}
\label{Fig1}
\end{figure}
As illustrated in Fig.~1 (see also Fig.~2 of the latter reference), this double-well potential is 
defined by
\begin{equation}
V(x) = \begin{cases} \infty & \text{if $x < 0$ or $x > a$} \\ 
V_0 & \text{if $(a-b)/2 <  x < (a+b)/2$} \\ 
V_L & \text{if $0 < x < (a-b)/2$} \\ 
V_R & \text{if $(a+b)/2 < x < a$}.
\end{cases}
\label{asy_doublewell_potential} 
\end{equation}
This potential describes two wells, each of width $w  \equiv (a-b)/2$, separated by a barrier of width $b$ and height $V_0$. 
The
``floor" level of each well is variable, but here we consider only the symmetric case, 
given by $V_L = V_R = 0$. A straightforward solution, valid for
$E< V_0$, is
\begin{align}
\psi_{I}(x) & = A\sin{qx} &  q  &\equiv \sqrt{2mE/\hbar^{2}} \notag \\
\psi_{II}(x) & = Be^{\kappa_2 x} + Ce^{-\kappa_2 x} & \kappa_2  &\equiv \sqrt{2m(V_{0}-E)/\hbar^{2}} \label{psi_regions} \notag \\
\psi_{III}(x) & = D\sin{q(a-x)} & \phantom{q} &\phantom{= \left( 2m(E - V_R)/\hbar^{2}\right)^{1/2},}
\end{align}
where the regions I, II, and III refer to $0 < x < w$, $w <  x < w+b$, and  $w+b < x < a$, respectively.
If we separately adopt suitable parameters for even and odd solutions (with respect to $x = a/2$), then matching the
\begin{figure}[h]
\includegraphics[width=6.0cm, angle=-90]{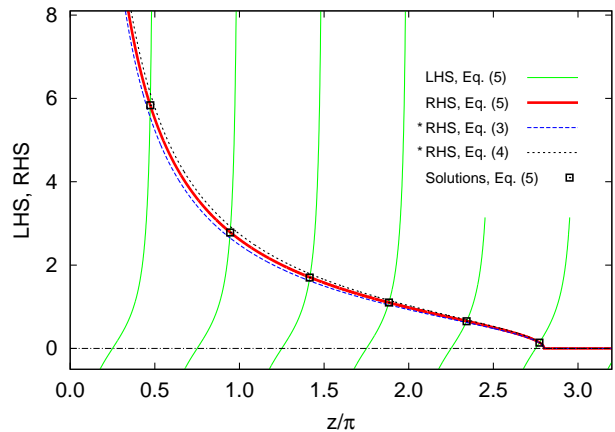}
\caption{Graphical solution of Eq.~(\ref{single_well}), for an example $z_0 = 2.8\pi$. The LHS is shown with
the thin solid (green) curves with the obvious characteristic branches of the $\tan$ function. The thick solid (red)
curve represents the RHS and lies centrally between two other curves. Intersections of LHS and RHS represent
solutions to Eq.~(\ref{single_well}), with the lowest energy solution just below $z_1 = \pi/2$; these solutions are
indicated by squares in the figure. The two thinner
curves bracketing the thick solid (red) curve are the RHSs of Eq.~(\ref{even}) [(blue) dashed curve slightly lower] and of
Eq.~(\ref{odd}) [(black) dotted curve slightly higher]. The asterisks on the labels indicate that the 
deviation from the central (red) curve has been exaggerated for clarity. The intersection of these curves with the 
thin green curve indicates a slight energy lowering and energy raising, respectively, with respect to the single well solution.
}
\label{Fig2}
\end{figure}
wave functions and their derivatives at the boundaries leads to
\begin{equation}
{\rm tan} (2z-\pi/2) = \left[\sqrt{\left({z_0 \over z}\right)^2 - 1}\right] {\rm tanh}\left({b \over w}\sqrt{z_0^2 - z^2}\right)
\label{even}
\end{equation}
for the even solution, and
\begin{equation}
{\rm tan} (2z-\pi/2) = \left[\sqrt{\left({z_0 \over z}\right)^2 - 1}\right] {\rm coth}\left({b \over w}\sqrt{z_0^2 - z^2}\right)
\label{odd}
\end{equation}
\noindent for the odd solution, where $z \equiv qw/2$ and $z_0 \equiv k_0 w/2$, with $k_0 \equiv \sqrt{2mV_0/\hbar^2}$.
If we use energy units $E_0 \equiv \hbar^2/(2mw^2)$, then $E/E_0 = 4z^2$. For a single well, i.e.~with $b/w \rightarrow \infty$, the hyperbolic functions are unity and the energy is given
by the solution of the simpler equation,
\begin{equation}
{\rm tan} (2z_1-\pi/2) = \sqrt{\left({z_0 \over z_1}\right)^2 - 1};
\label{single_well}
\end{equation}
we use $z_1$ do denote this single well solution, and presume that it is obtained numerically or on a calculator by iteration.
The energy corresponding to this level will be denoted $E_1 = 4z_1^2 E_0$.
It is not hard to see that Eq.~(\ref{even}) results in a slightly lower energy solution (compared to the energy $E_1$,
the solution of Eq.~(\ref{single_well})), while Eq.~(\ref{odd}) results in a slightly higher energy solution. Mathematically this is
because the hyperbolic tangent is always slightly less than unity while the hyperbolic cotangent is always slightly greater than
unity. Physically this corresponds to the bonding and anti-bonding solutions to a particle which is given freedom to move in
two basins (i.e.~an electron free to roam among two atoms in a molecule). The situation is illustrated graphically in 
Fig.~\ref{Fig2}; the left-hand-sides
(LHSs) are the same in all three equations, and are indicated by the solid (green) curves. The right-hand-side (RHS) 
of Eq.~(\ref{single_well}) is indicated by the thick solid (red) curve that lies in between the two curves representing the RHSs of
Eq.~(\ref{even}) (in blue, below) and Eq.~(\ref{odd}) (in black, above), with a slightly lower and higher energy, respectively. This
fine splitting of an otherwise degenerate level is what is expected for a significant barrier between the two wells. As stated earlier,
generically one expects that coupling $N$ wells will result in a splitting into $N$ energies.

Let us focus on the most tightly bound, lowest, energy level. Then the argument in the hyperbolic functions will be
very close to unity; expanding to first order results in
\begin{equation}
{\rm tan} (2z-\pi/2) = \sqrt{\left({z_0 \over z}\right)^2 - 1} \left[ 1 \mp 2 \ {\rm exp}\left\{ -2{b \over w}\sqrt{z_0^2 - z^2} \right\} \right],
\label{expand}
\end{equation}
where the minus [plus] sign results from Eq.~(\ref{even}) [\ref{odd}], and the exponential correction is expected to be very small.
We thus look for solutions
\begin{equation}
z_{e,o} = z_1 \mp \rho,
\label{expandb}
\end{equation}
where, as mentioned above, $z_1$ is presumed known (and somewhat less than $\pi/2$), and
the subscript `$e$' (`$o$') corresponds to the even (odd) solution. Inserting Eq.~(\ref{expandb}) into
Eq.~(\ref{expand}), and expanding everywhere to first order in $\rho$ results in
\begin{equation}
\rho = 2\delta z_0 {1 - \delta^2 \over {1 + 2 z_0\sqrt{1 - \delta^2}}} {\rm exp}\left\{ - 2{b \over w}z_0 \sqrt{1 - \delta^2}\right\},
\label{rho}
\end{equation}
where $\delta$ is the single-well energy level, $\delta \equiv z_1/z_0 = \sqrt{E_1/V_0}$, determined in advance.

If we use the values from Ref.~[\onlinecite{dauphinee15}], i.e.~$V_0 = 500 {\pi^2 \hbar^2 / (2 m a^2)} = 500 \pi^2 E_0 w^2/a^2
= 80 \pi^2 E_0$ for $w/a = 2/5$, then $z_0 = \pi \sqrt{80}$. We can solve Eq.~(\ref{single_well}) on a calculator, and we obtain
$\delta \approx 0.108$. Plugging this into Eq.~(\ref{rho}) we find $\rho \approx 1.78 \times 10^{-7}$. 

The ``toy model'' here is a two-state system, as in the Feynman example, but with a wave function describing the
particle to be in the left well ($\psi_L$ ) and a wave function for the particle in the right well ($\psi_R$).  The tunneling amplitude $t$
mentioned in the Introduction and
defined in Eq.~(11) of Ref.~[\onlinecite{dauphinee15}] as the matrix element for tunneling from the left well into the right well (or
vice-versa), is defined by the correspondence between the energy there, $E = E_1 - t$, and the energy here, 
$E = 4 z_1^2 E_0  -8E_0\rho z_1$. More explicitly, we repeat here Eq.~(11) from Ref.~[\onlinecite{dauphinee15}]:
\begin{equation}
\begin{aligned}
H\psi_L &= E_1 \psi_L - t\psi_R, \\
H\psi_R &= E_1 \psi_R - t\psi_L,
\end{aligned}
\label{feynman}
\end{equation}
which describes the coupling between the two states through the parameter $t$. Comparing to the expression
above Eq.~(\ref{feynman}) shows that $t \equiv 8E_0 \rho z_1$. Therefore,
\begin{equation}
t = 16E_0 {z}_1^2 \frac{1 -\delta^2}{1+2 z_0 \sqrt{1 -\delta^2}} 
e^{-2\frac{b}{w} z_0 \sqrt{1 - \delta^2}};
\label{eq:t_doub}
\end{equation}
with the parameters used above, we obtain 
\begin{equation}
t \approx 1.08 \times 10^{-6}E_0,
\label{thop}
\end{equation}
and in the units of Ref.~[\onlinecite{dauphinee15}], we have
\begin{equation}
t \approx 6.84 \times 10^{-7} {\pi^2 \hbar^2 \over 2 ma^2},
\label{thopb}
\end{equation}
which is precisely what was obtained there through a fit to the numerical data.

In summary, we have obtained the toy model parameter $t$, which describes the transition amplitude for the
particle to tunnel from the left side of the double-well to the right side (or vice-versa), in terms of characteristics
of the microscopic model and parameters involving the single well. In Ref.~[\onlinecite{dauphinee15}] a qualitative
estimate was provided, based on a WKB approximation. Here we have improved considerably on this estimate, and
now have a quantitatively accurate
correspondence between the ``microscopic'' double-well potential and the two-state system. 

In Appendix~A we briefly discuss another example of a simple double-well potential, where a similar correspondence
with a ``toy" model is achieved. This model has identical single-well characteristics to that of the Kronig-Penney model
to be discussed in the next section; for the Kronig-Penney model the ``toy'' model is the tight-binding formulation of one
of the bands present in this model.

\section{Kronig-Penney Model}
\label{sec:kp}
\begin{widetext}
Fig. 3
\begin{figure} [h]
\centering
\begin{tikzpicture}
	\draw [ultra thick, <->] (-5,0) -- (5.5,0);
	\draw [ultra thick, ->]  (0,0) -- (0,4);
	\draw [ultra thick, -]   (-1,0) -- (-1,3) -- (0,3);
	
	\draw [ultra thick, -]   (2,0) -- (2,3) -- (3,3) -- (3,0);
	\draw [ultra thick, -]   (-4,0) -- (-4,3) -- (-3,3) -- (-3,0);
	\draw [ultra thick, -]   (5,0) -- (5,3) -- (5.5,3);
	
	\node [above right] at (0,4) {$V(x)$};
	\node [right] at (0,3) {$V_0$};
	\node [right] at (5.5,0) {$x$};
	\node [below] at (0,0) {$0$};
	\node [below] at (3,0) {$w+b$};
	\node [below] at (2,-.1) {$w$};
	\node [below] at (-1.1,0) {$-b$};

\end{tikzpicture}
\caption{A pictorial representation of the periodic potential in the Kronig-Penney model, illustrating wells of depth
$V_0$ and width $w$ separated from one another by barriers of width $b$.} 
\label{fig:kpmodel}
\end{figure}
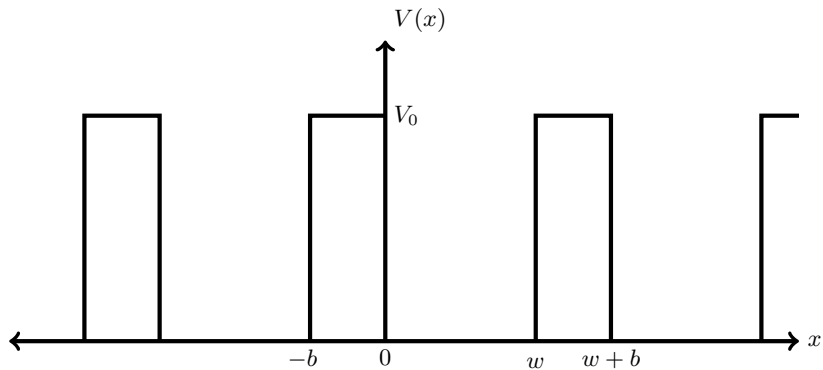
\end{widetext}

The one-dimensional Kronig-Penney model\cite{kronig31} consists of an electron moving in a periodic potential as depicted in
Fig.~\ref{fig:kpmodel}, with alternating wells of width $w$ and barriers of width $b$ and height $V_0$. 
The analytical solution for the energy levels ($E < V_0$) is well known; the implicit equation for the energy is
\begin{widetext}
\begin{equation}
\cos{(k\ell)} = \cos{(q w)} \cosh{(\kappa_2 b)} + \frac{\kappa_2^2 - q^2}{2q \kappa_2} \sin{(q w)} \sinh{(\kappa_2 b)},
\label{eq:kpanalsol1Dsec}
\end{equation}
\end{widetext}
where $\ell = w + b$ is the unit cell length, and $q = \sqrt{2mE/\hbar^2}$ and $\kappa_2 = \sqrt{2m(V_0-E)/\hbar^2}$. For each
wave vector $k$, with values $-\pi < k\ell \le \pi$, one needs to solve this equation for $E(k)$.
As is well known, the periodicity in the problem
gives rise to a series of energy bands as a function of wave vector $k$, with each band separated by an energy gap. In the case where
the wells illustrated in Fig.~\ref{fig:kpmodel} are deep, any single well, taken in isolation, would consist of a number of
different energy levels corresponding to states that are bound within each well. As already stated, when $N$ of these wells are
coupled through barriers, each of these energy broadens into $N$ states, forming bands. Numerical solutions to this and
other periodic models with different potential shapes are given in Refs.~[\onlinecite{pavelich15}]~and~[\onlinecite{pavelich16}].

The tight-binding limit tends to focus on one of these bands, and is used to describe the dispersion, $E(k)$ for this band. General 
considerations\cite{ashcroft76} in the tight-binding limit in one dimension lead to a dispersion of the form
\begin{equation}
E(k) = E_b - 2t_1\cos(k\ell) - 2t_2\cos(2k\ell) - 2t_3\cos(3k\ell) - \ldots
\label{eq:tightbindingeps}
\end{equation}
The usual interpretation of such a dispersion is that each additional term corresponds to tunneling of an electron
from a well to a further neighboring well. In other words, while $t_1$ represents a tunneling amplitude for an
electron to tunnel through one of the barriers in Fig.~\ref{fig:kpmodel}, $t_2$ represents a tunneling amplitude for the
electron to tunnel through two of the barriers, and end up (directly) two unit cells away from its initial location. Given
that the electron wavefunctions are exponentially decaying in the barrier regions, it should be clear 
that $\abs{t_1} \gg \abs{t_2} \gg \abs{t_3} \gg \ldots$ in this limit. In what follows we will first focus on nearest-neighbor tunneling
amplitudes only, i.e. we will obtain from Eq.~(\ref{eq:kpanalsol1Dsec}), an explicit expression for $t_1$.

\begin{figure}[h]
\includegraphics[width=6.0cm,angle=-90]{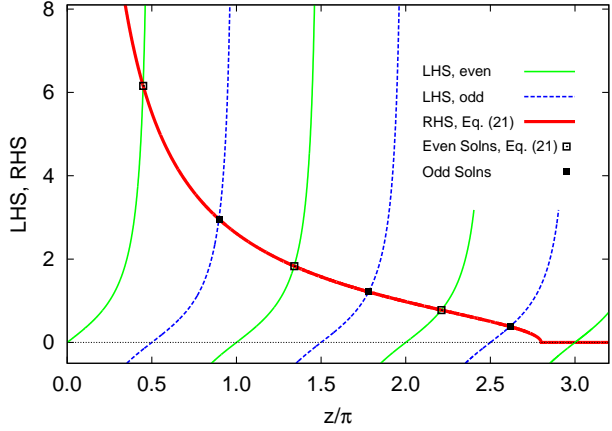}
\caption{Graphical solution of Eq.~(\ref{eq:tanz}), for an example $z_0 = 2.8\pi$. The LHS is shown with
the thin sold (green) curves, with the obvious characteristic branches of the $\tan$ function. The thicker solid (red)
curve represents the RHS. The intersections of these two curves represent the {\it even} bound states and are indicated by
open squares; the lowest energy solution is just below $\tilde{z}_1 = \pi/2$. For completeness we have also drawn the LHS and
RHS for the odd bound states. The LHS is given by ${\rm tan}(\tilde{z}_1 - \pi/2)$ and is shown with thin dashed (blue)
curves. The RHS is the same solid (red) curve as for the even states. Their intersections are indicated by filled squares.
For our tight binding solutions we will focus on the lowest energy (even) bound state, i.e.~the point with 
$\tilde{z}_1 {{ \atop <} \atop {\sim \atop }}
\pi/2$.
}
\label{Fig4}
\end{figure}

When $V_0$ or $b$ is suitably large, one can rewrite Eq.~(\ref{eq:kpanalsol1Dsec}) to obtain\cite{remark1}
\begin{widetext}
\begin{equation}
\left( \cos{\frac{qw}{2}} - \frac{q}{\kappa_2}\sin{\frac{qw}{2}}\right) 
\left( \cos{\frac{qw}{2}} + \frac{\kappa_2}{q}\sin{\frac{qw}{2}} \right) = \eta_1(k) + \eta_2
\label{eq:nu}
\end{equation}
where
\begin{eqnarray}
\eta_1(k) &=& 2 \, e^{-\kappa_2 b} \cos{(k\ell)} \nonumber \\
\eta_2\phantom{aa} &=& -e^{-2\kappa_2 b} \left( \cos{\frac{qw}{2}} - \frac{\kappa_2}{q} \sin{\frac{qw}{2}}\right) 
\left( \cos{\frac{qw}{2}} +  \frac{q}{\kappa_2} \sin{\frac{qw}{2}} \right).
\label{eq:nu2}
\end{eqnarray}
\end{widetext}
Written in this way, it is easy to see that when there is no coupling between the wells
(e.g.~put $b \rightarrow \infty$) and therefore both $\eta_1(k) \rightarrow 0$ and $\eta_2 \rightarrow 0$, 
then the vanishing of the first (second) factor on the LHS of Eq.~(\ref{eq:nu})
corresponds to determining the energy for the even (odd) bound states in the single well. It is convenient to define dimensionless
variables as before, specifically $z \equiv q w/2$ and $z_0 \equiv k_0 w/2$, where $k_0 \equiv \sqrt{2mV_0/\hbar^2}$. Then
Eq.~(\ref{eq:nu}) reads
\begin{widetext}
\begin{equation}
\left( \cos{z} - \frac{z}{\sqrt{z_0^2 - z^2}} \sin{z} \right) 
\left( \cos{z} + \frac{\sqrt{z_0^2 - z^2}}{z}\sin{z} \right) = \eta_1(k) + \eta_2,
\label{eq:nuz}
\end{equation}
where we have used $\kappa_2 w/2 = \sqrt{z_0^2 - z^2}$ and now
\begin{equation}
\eta_1(k) = 2 e^{-2 \frac{b}{w} \sqrt{z_0^2 - \tilde{z}_1^2}} \cos{k \ell}
\label{eq:nuz1}
\end{equation}
and
\begin{equation}
\eta_2 = e^{-4 \frac{b}{w} \sqrt{z_0^2 - \tilde{z}_1^2}} \left( \cos{z} - \frac{\sqrt{z_0^2 - z^2}}{z} \sin{z} \right) 
\left( \cos{z} + \frac{z}{\sqrt{z_0^2 - z^2}}\sin{z} \right).
\label{eq:nuz2}
\end{equation}
\end{widetext}
Eq.~(\ref{eq:nuz}) is still exact; now we can imagine the scenario where $b/w$ is very large, and hence the RHS
of this equation is very small.
The zeroth order solution for the even
bound state is given by setting the first factor on the LHS of Eq.~(\ref{eq:nuz}) to zero,
\begin{equation}
\cos{\tilde{z}_1} - \frac{\tilde{z}_1}{\sqrt{z_0^2 - \tilde{z}_1^2}} \sin{\tilde{z}_1} = 0,
\label{eq:cosdelta}
\end{equation}
and this determines the zeroth order solution, $\tilde{z}_1$. The equation to determine $\tilde{z}_1$ can be written as
\begin{equation}
\tan{\tilde{z}_1} = \sqrt{\left( \frac{z_0}{\tilde{z}_1}\right)^2 - 1},
\label{eq:tanz}
\end{equation}
which is the equation that determines the bound state energy for a particle in a {\it single} well of width $w$ and depth $V_0$.
The solution is shown graphically in Fig.~\ref{Fig4}.
An actual number for $\tilde{z}_1$, slightly less than $\pi/2$, is readily obtained numerically or on a calculator.\cite{remark2}
\begin{widetext}
Fig. 5
\begin{figure}[h]
\includegraphics[width=5.1cm,angle=-90]{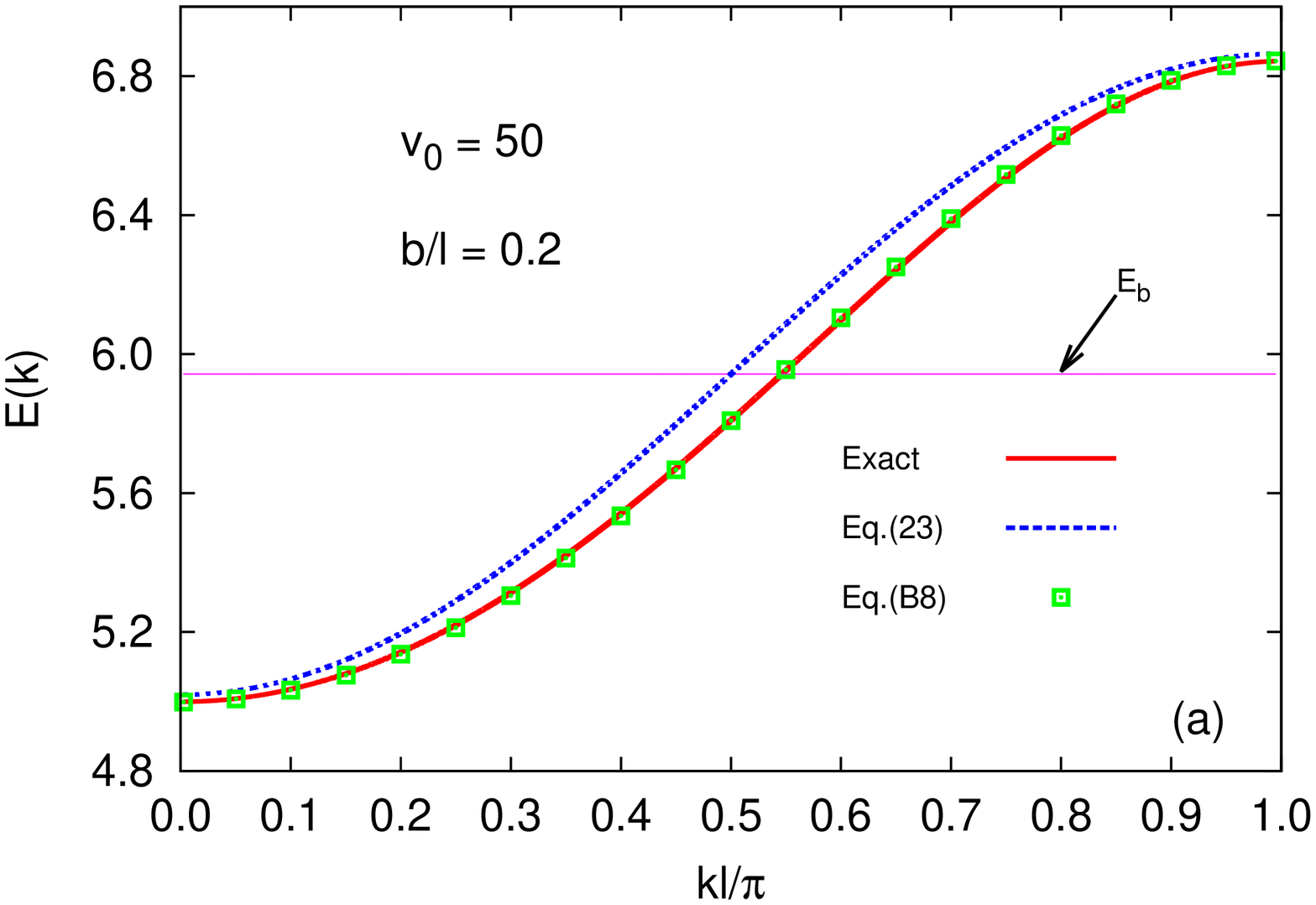}
\includegraphics[width=5.1cm,angle=-90]{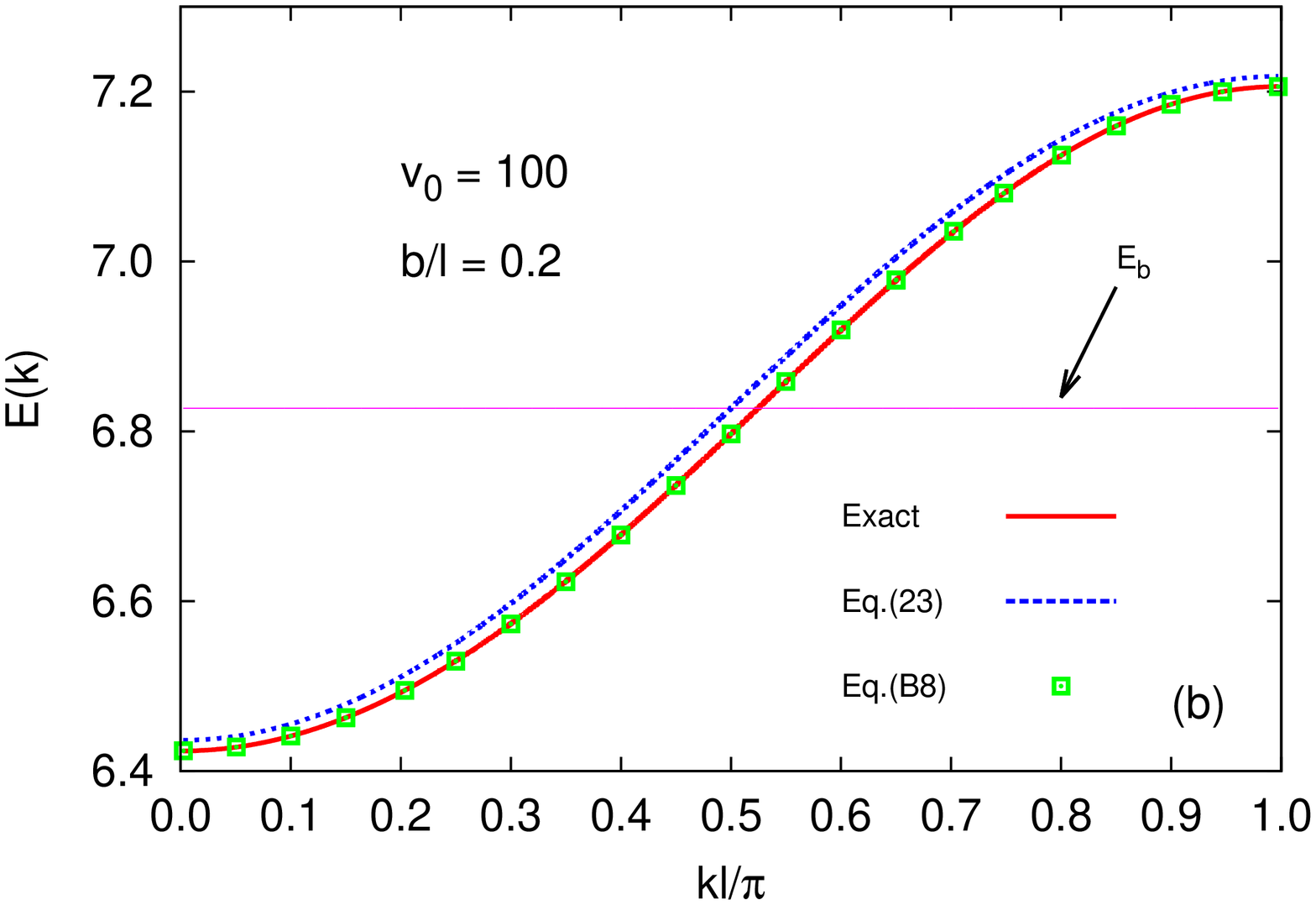}
\includegraphics[width=5.1cm,angle=-90]{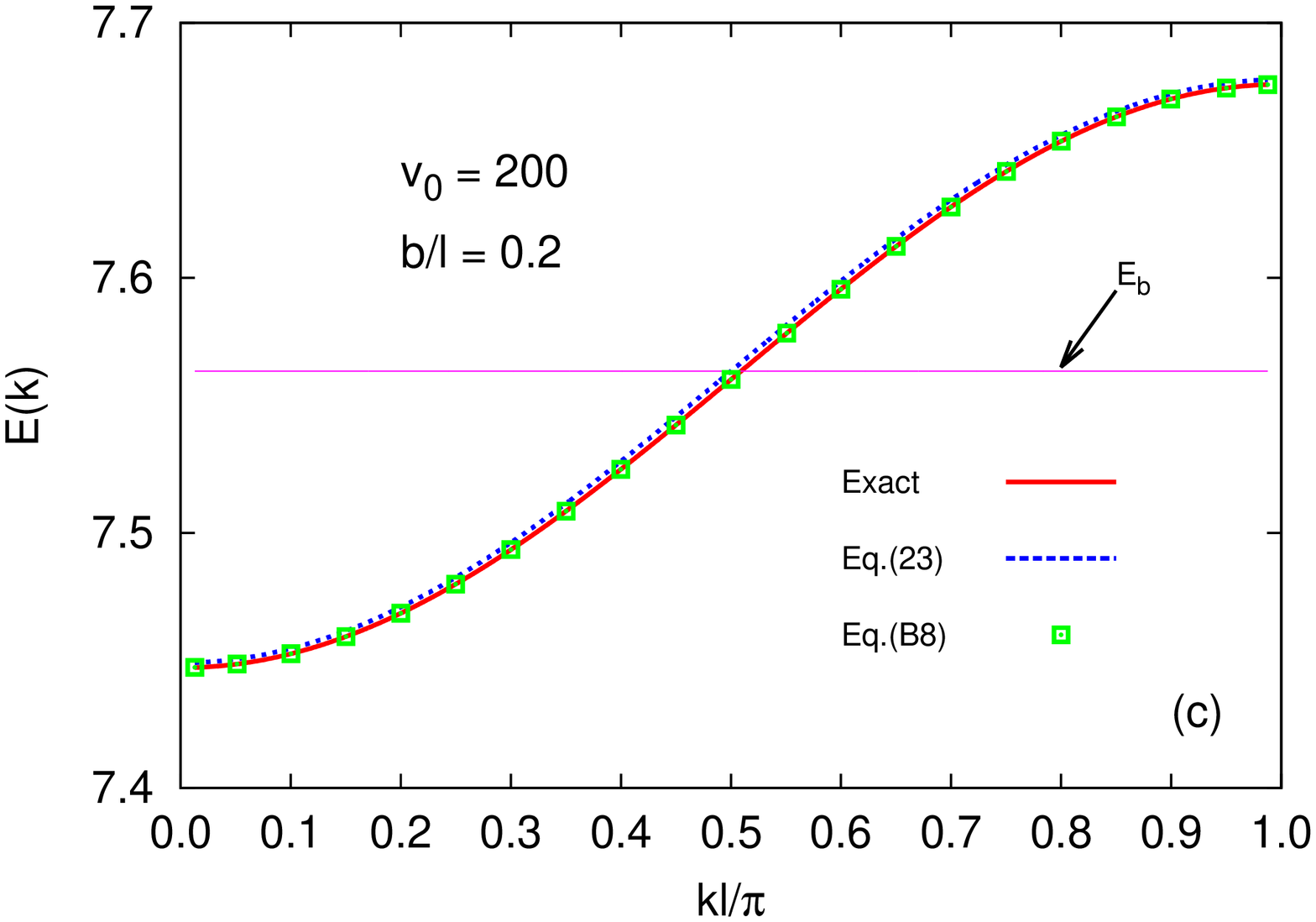}
\includegraphics[width=5.1cm,angle=-90]{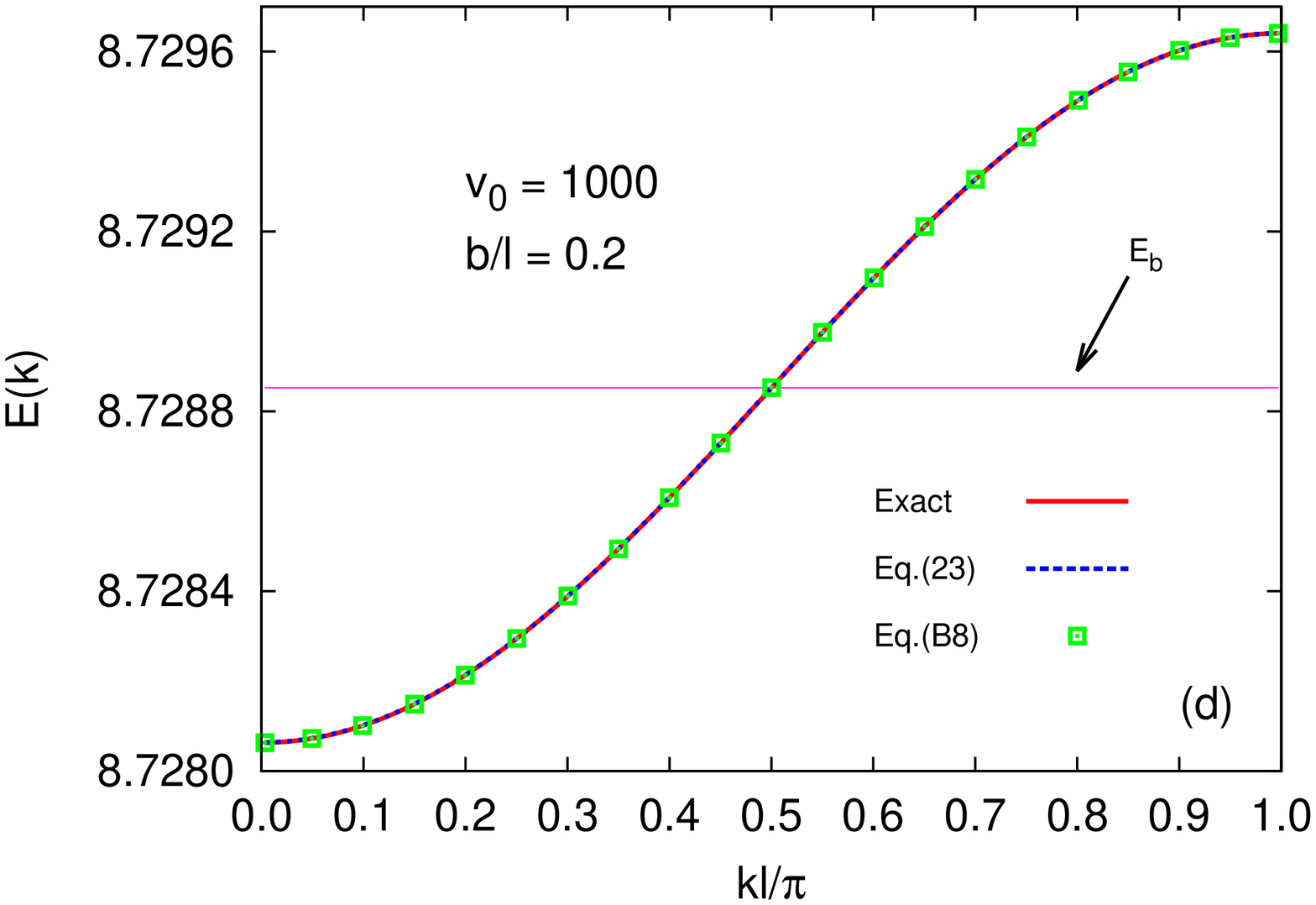}
\caption{Comparison of various approximations with the exact result (solid (red) curve). All results are for $b/\ell = 0.2$ and
are for (a) $v_0 = 50$, (b) $v_0 = 100$, (c) $v_0 = 200$, and (d) $v_0 = 1000$. In (a), for example, the first-order result is given 
by Eq.~(\ref{eq:tightbindinganal}) (shown as the dashed (blue) curve) with the
exact result determined numerically from Eq.~(\ref{eq:kpanalsol1Dsec}) (shown as the solid (red) curve). 
Note that there remains a significant discrepancy. The zeroth-order result, a constant given by the first term only in
Eq.~(\ref{eq:tightbindinganal}), is shown as the horizontal (pink) line at $E/E_0 \approx 5.94$.
With further second-order corrections that arise from $\eta_2$ and the nonlinear nature
of Eq.~(\ref{eq:nuz}), we also show the results from Eq.~(\ref{ener_2ndorder}) in Appendix~B. This result now agrees very well with the
exact result, and includes terms corresponding to $-2t_2 {\rm cos}(2k\ell)$ in Eq.~(\ref{eq:tightbindingeps}).
Notice that for the last case, all curves and points are
essentially in agreement. This is even more impressive given the reduction in the scale of the bandwidth. 
}
\label{Fig5}
\end{figure}
\begin{figure}[h]
\includegraphics[width=5.0cm,angle=-90]{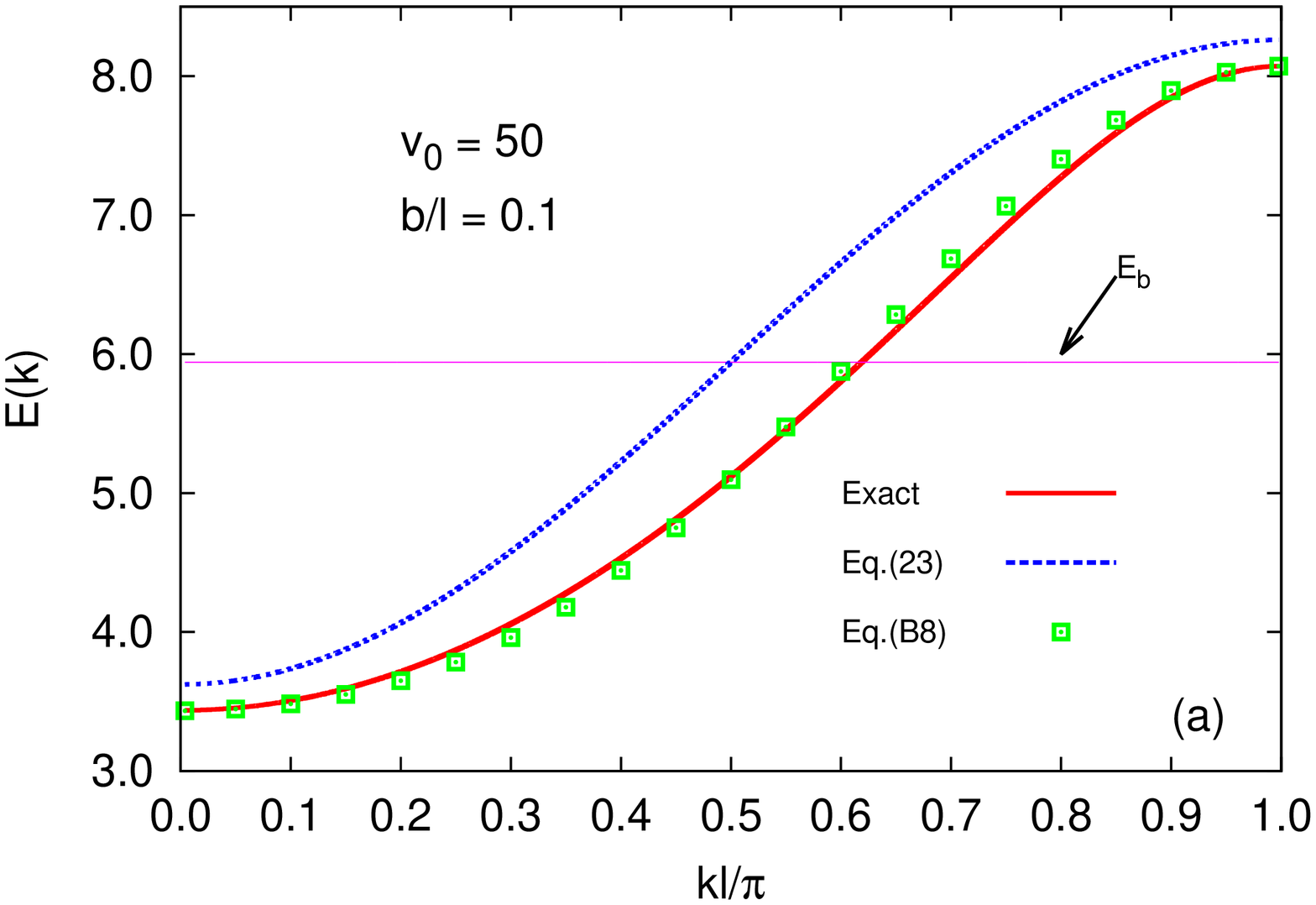}
\includegraphics[width=5.0cm,angle=-90]{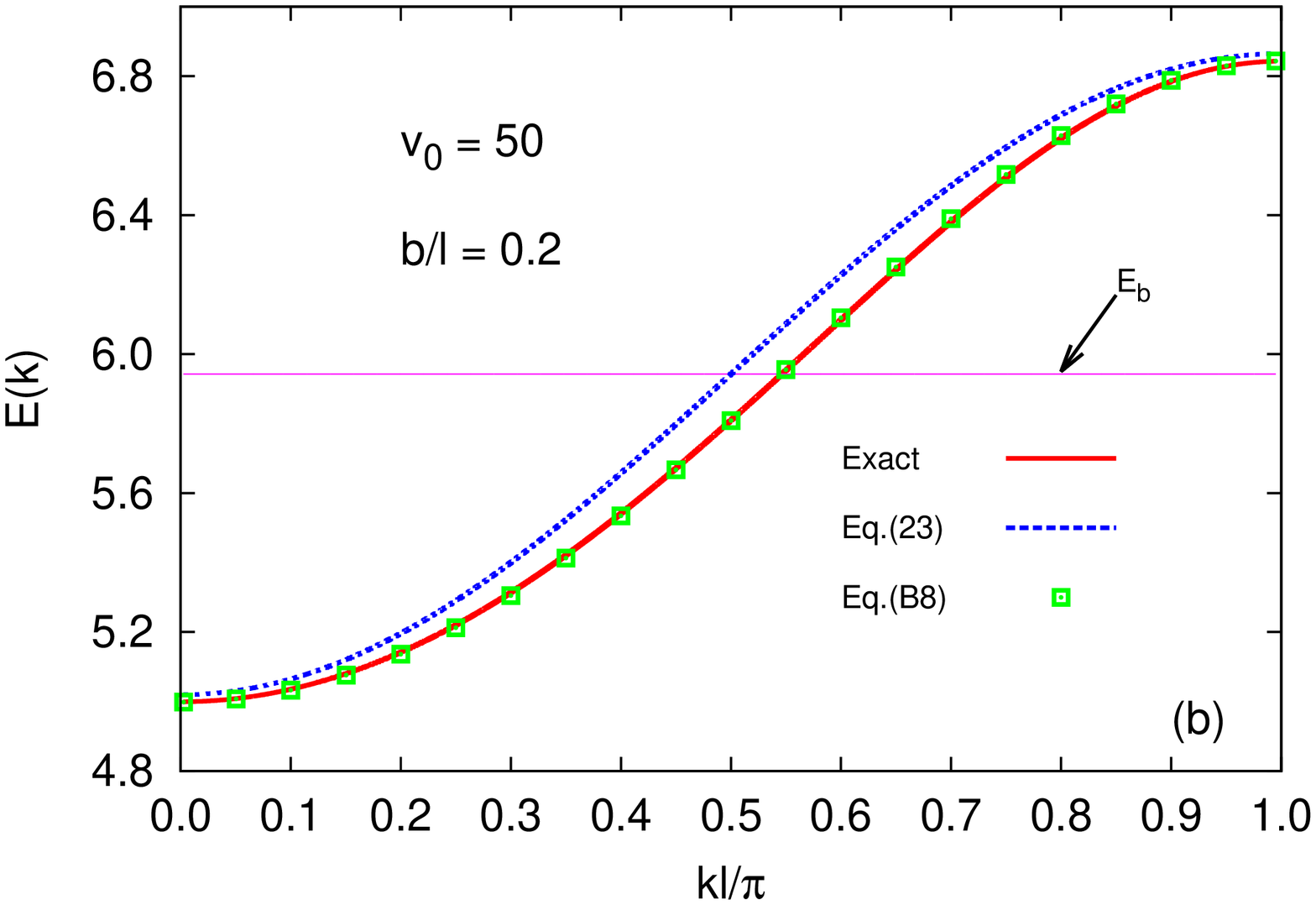}
\includegraphics[width=5.0cm,angle=-90]{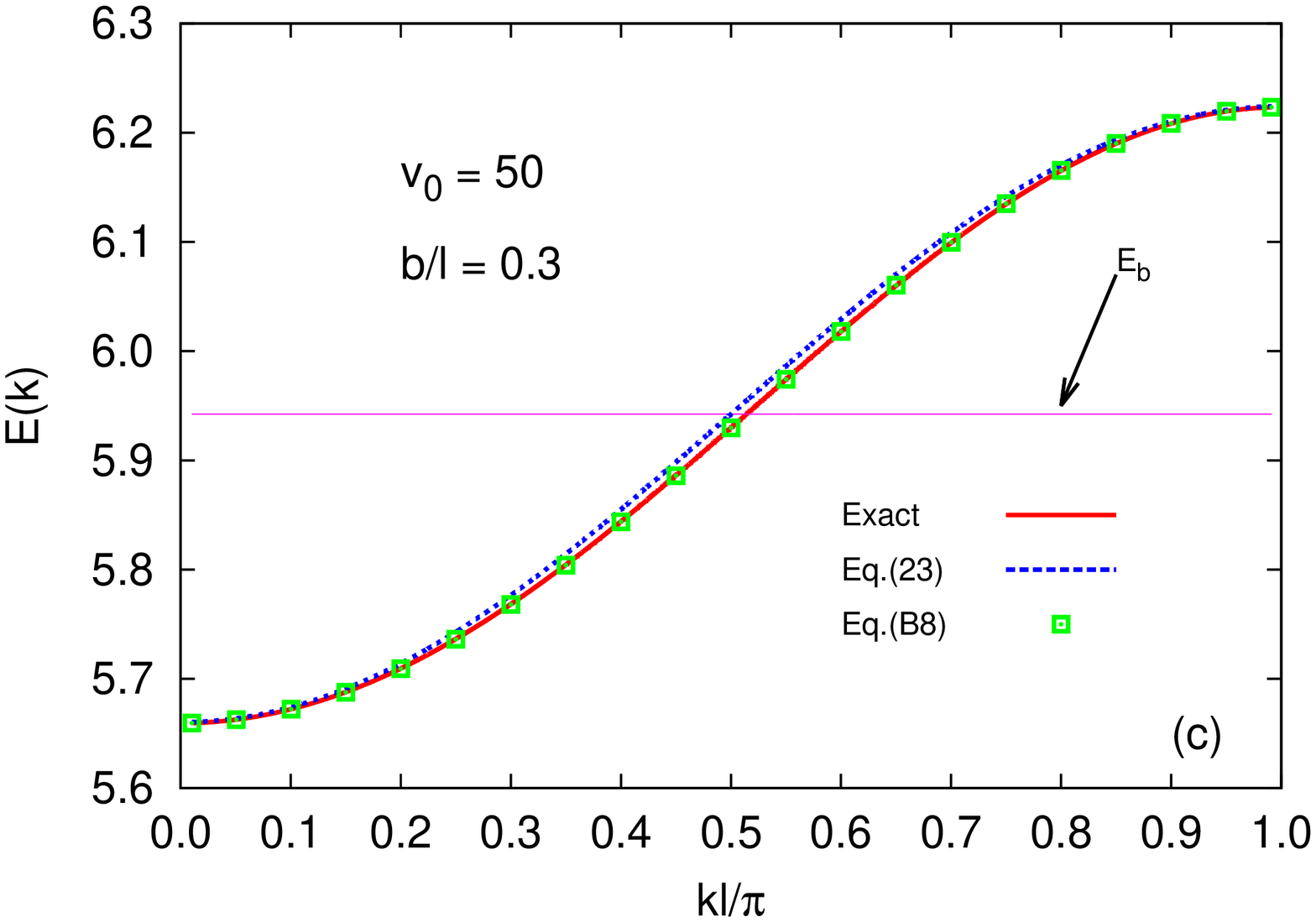}
\includegraphics[width=5.0cm,angle=-90]{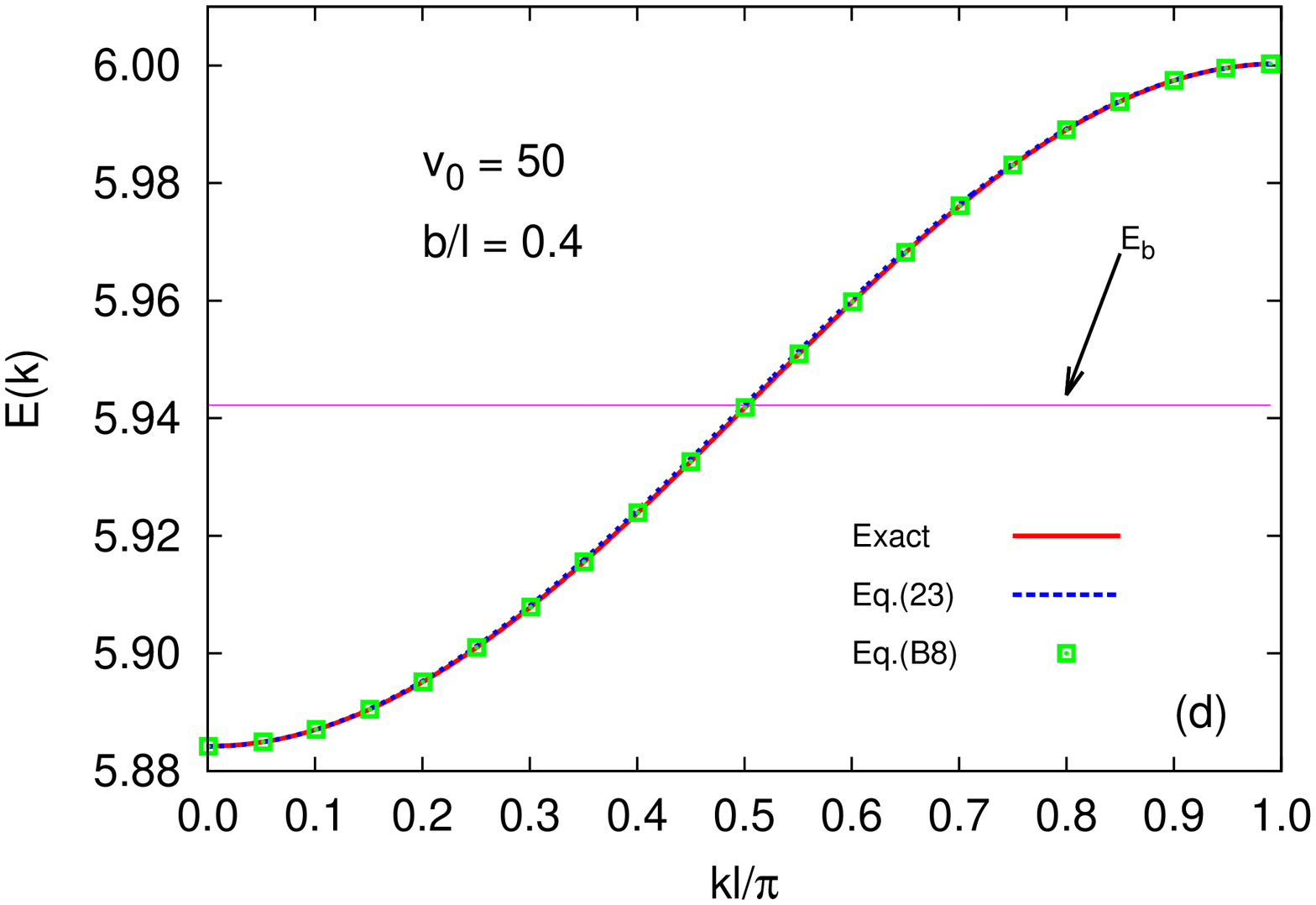}
\caption{As in Fig.~\ref{Fig5}, a comparison of various approximations with the exact result (solid (red) curve)
for a variety of barrier widths, all with $v_0 \equiv V_0/E_0 = 50$. We use barrier widths of
(a) $b/\ell = 0.1$, (b) $b/\ell = 0.2$, (c) $b/\ell = 0.3$ and (d) $b/\ell = 0.4$. By examining the vertical scales, a 
clear progression towards more tightly bound
wells is evident. Moreover, the first case considered shows not perfect agreement, even when $2^\text{nd}$ order corrections are included.
As $b/\ell$ increases, agreement quickly improves, especially when account is made of the steady reduction in the scale 
of the bandwidth. For all cases, the horizontal (pink)
line corresponds to the bound state energy for a single well and corresponds to the constant given by the first term only in
Eq.~(\ref{eq:tightbindinganal}).
The approximate result with the $1^\text{st}$ order correction only, the full
Eq.~(\ref{eq:tightbindinganal}), is shown as a dashed (blue) curve, while the approximate results including $2^\text{nd}$ order
corrections from Appendix B are given by the square (green) symbols.
}
\label{Fig6}
\end{figure}
\end{widetext}

\begin{figure}[h]
\includegraphics[width=5.0cm,angle=-90]{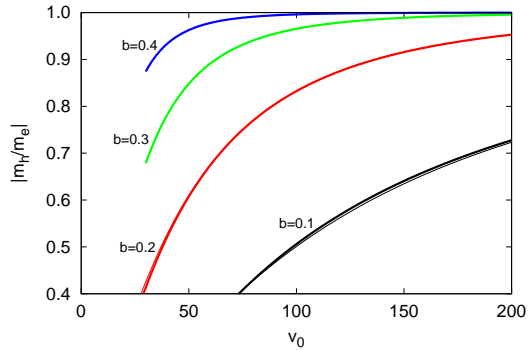}
\caption{Absolute value of the effective mass ratio,  $|m_h/m_e|$ vs. $v_0 \equiv V_0/E_0$ for the various values of
barrier width $b$, as indicated. Here $m_e^{-1} \equiv {\partial^2 E(k)/E_0 \over \partial (k\ell)^2}$ at $k = 0$ and
$m_h^{-1} \equiv {\partial^2 E(k)/E_0 \over \partial (k\ell)^2}$ at $k = \pi/\ell$. The thick curves are the exact result,
determined from Eq.~(\ref{eq:kpanalsol1Dsec}), while the thinner curves are determined from the tight-binding
parametrization of Eq.~(\ref{ener_2ndorder}). These latter curves are barely visible over almost the entire parameter regime
shown, indicating that when the 2nd order corrections considered in Appendix B are included the derived tight-binding
parameters are very accurate.
}
\label{Fig7}
\end{figure}

A more accurate solution to Eq.~(\ref{eq:nuz}) can be obtained to $1^\text{st}$ order in $\eta_1(k)$ by writing 
$z = \tilde{z}_1\left[ 1 + \tilde{\rho}(k)\right]$, and expanding
that equation to $1^\text{st}$ order in $\tilde{\rho}(k)$. After some algebra we obtain
\begin{equation}
\tilde{\rho}(k) = -\frac{2}{z_0^2} 
\frac{(z_0^2 - \tilde{z}_1^2)}{(1 + \sqrt{z_0^2 - \tilde{z}_1^2})} e^{- \frac{2b}{w} \sqrt{z_0^2 - \tilde{z}_1^2}} \cos{k \ell}.
\label{1st_order}
\end{equation}
Note that we have ignored $\eta_2$, as that factor is exponentially suppressed with respect to $\eta_1(k)$.
Using the energy scale $E_0 = \hbar^2/ (2mw^2)$ as before, we find for the energy to $1^\text{st}$
order in $\eta_1(k)$,
\begin{equation}
\frac{E}{E_0} = 4\tilde{z}_1^2 - \left({4\tilde{z}_1 \over z_0}\right)^2 
\frac{z_0^2 - \tilde{z}_1^2}{1+\sqrt{z_0^2 - \tilde{z}_1^2}} e^{-2 \frac{b}{w} \sqrt{z_0^2 - \tilde{z}_1^2}} \cos{k \ell}
\label{eq:tightbindinganal}
\end{equation}
which has the functional form of nearest-neighbor tight-binding (see Eq.~(\ref{eq:tightbindingeps})). A comparison of this
result with the exact result is shown in Fig.~\ref{Fig5} as a function of wave vector for particular values of $V_0$ and $b/\ell$.

If we first focus on case (a) in Fig.~5, the 
first-order result has significant disagreement with the exact result. It is indeed true that improved agreement is readily
attained by using deeper wells, as is clear from the progression through (b)--(d). 
However, in Appendix B we sketch a more involved derivation to $2^\text{nd}$ order in $e^{-x}$, where
$x \equiv 2 \frac{b}{w} \sqrt{z_0^2 - \tilde{z}_1^2}$---see Eq.~(\ref{ener_2ndorder}). The resulting expression is 
plotted in Fig.~\ref{Fig5} as square points; these lie essentially on top of the exact results for all four cases, even for case (a).
In going to $2^\text{nd}$ order in $e^{-x}$ we automatically generate constant corrections to the energy, as well as corrections
with dispersion, namely those corresponding to $\cos{2k\ell}$. The need to include these terms, as is the case for the result
in the first two cases of Fig.~\ref{Fig5}, is often taken to be indicative of a significant tunneling process directly to the
next-nearest neighbor. While this is in part
correct, we note that in Appendix A we examine the double-well potential, and observe there the need for corrections of $2^\text{nd}$
order, i.e.~terms in the energy proportional to $e^{-2x}$. For the double-well potential, however, there are no 
next-nearest-neighbor wells, as there are only two in total! It is a difficult problem to disentangle contributions to $2^\text{nd}$ order
from next-nearest-neighbor tunneling  and contributions arising from the inherent non-linear nature of the equations; at this point
we simply caution that all these contributions are not entirely due to direct next-nearest-neighbor tunneling.


For completeness, in Fig.~\ref{Fig6} we fix the well depth to be $v_0 \equiv V_0/E_0 =  50$ and show the 
dispersions for a variety of different barrier widths.
The trends are the same in the two cases, except that the $2^\text{nd}$ order result is not very accurate for the least tightly
bound case considered (a). With increasing barrier width, however, as in Fig.~\ref{Fig5}, both the $2^\text{nd}$-order  and the
$1^\text{st}$-order results become increasingly accurate. Note the change in scale as $v_0$ and $b/\ell$ increase in Fig.~\ref{Fig5} 
and Fig.~\ref{Fig6}, respectively; in both cases the results approach the single well result while a well-defined dispersion remains.


In both Fig.~\ref{Fig5} and Fig.~\ref{Fig6} it should be clear that for the more strongly coupled wells (e.g.~(a) and (b) in particular)
a significant amount of electron-hole asymmetry is present. In Fig.~\ref{Fig7} the effective mass ratio, $|m_h/m_e|$ is shown
for various values of the barrier thickness as a function of well depth. Here,
the electron mass, $m_e$ is defined in the usual way (see Fig.~\ref{Fig7}) through the curvature at $k=0$ and similarly for the hole mass,
$m_h$. As discussed in Ref.~[\onlinecite{pavelich15}] an asymmetry is expected on general grounds since holes are by definition
closer to the top of the barriers than electrons. They should therefore have lower masses for this reason alone, and this is
reflected in the results of Fig.~\ref{Fig7}, where all the ratios are lower than unity. The thicker curves are from the exact calculations
while the thinner curves (not visible for most of the parameter space shown) are readily determined from the tight-binding
parametrization of Eq.~(\ref{ener_2ndorder}). These are fairly accurate when the higher-order correction considered
in Appendix B is included.

\section{Summary}

We have succeeded in deriving the effective model for the periodic potential first used to model a solid, the so-called Kronig-Penney
model, consisting of a series of wells and barriers. We first started with a double-well, and we were able to achieve very high accuracy
by exploiting the tightly-bound limit, where two neighboring wells are well separated. This ensures that the tunneling amplitude
between the two wells is very small, and one can essentially use perturbation theory with respect to this ``atomic limit.'' In particular we
considered the double-well of Ref. [\onlinecite{dauphinee15}], and we were able to achieve {\it quantitative} agreement with the
numerical results obtained there.

The generalization of this process to an infinite array of wells and barriers is straightforward. However, incorporating tunneling
to $1^\text{st}$ order (meaning terms of order $e^{-x}$, where $x \propto b\sqrt{V_0}$) provided only a {\it qualitative} agreement with the
exact result (this would become quantitative for sufficiently large $b/\ell$ and/or $V_0$). When including terms of $2^\text{nd}$ order
in $e^{-x}$, very good quantitative agreement was achieved, even for moderate well depths. Terms of order $e^{-2x}$ would
necessarily be accompanied by dispersive terms like $\cos{(2k\ell)}$, which are generally associated with next-nearest-neighbor
tunneling, i.e.~tunneling across two barriers. By comparison with results of a simple double-well, where terms of order $e^{-2x}$
also contribute to the energy, we were able to show that terms of this order were not exclusively associated with such longer-range
tunneling. Instead, inherent non-linearity of the equations governing the electronic energy dispersion will naturally
give rise to such terms, even in the absence of next-nearest-neighbor tunneling. 

It is useful to map the complete microscopic double-well problem onto the two-state system that is often used to describe this
problem in simplified terms. Similarly, it is useful to map the microscopic problem of an infinite array of wells onto a simplified
model---this is the tight-binding description. We have carried out such a mapping, with no ``fitting'' involved and we have illustrated the
accuracy as well as the limitations of such a mapping.

\section{ACKNOWLEDGEMENTS}
This work was supported in part by the Natural Sciences and Engineering
Research Council of Canada (NSERC) and originated in work originally funded by a 
University of Alberta Teaching and Learning Enhancement Fund (TLEF) grant, for which we are grateful.

\appendix
 
\section{A simple double-well potential}

\begin{widetext}
Fig. A1
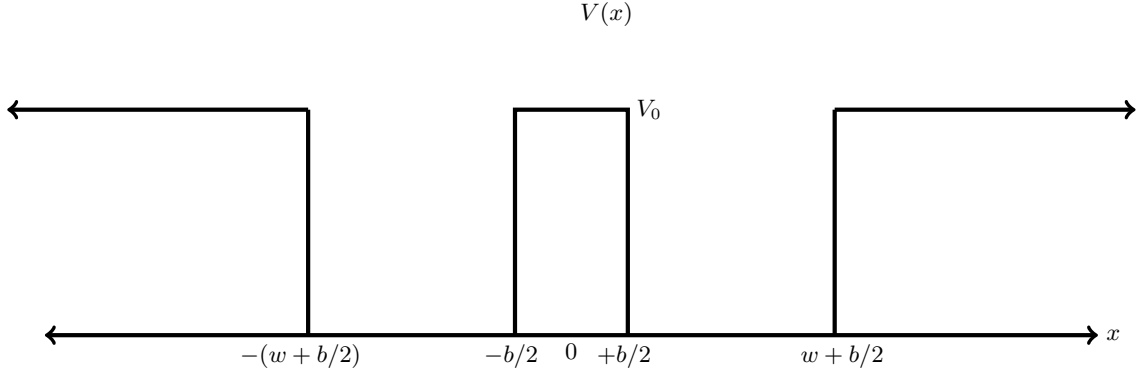
\begin{figure} [h]
\centering
\begin{tikzpicture}
	\draw [ultra thick, <->] (-7,0) -- (7,0);
	
	\draw [ultra thick, -]   (-0.75,0) -- (-0.75,3) -- (0.75,3) -- (0.75,0);
	\draw [ultra thick, -]   (3.5,0) -- (3.5,3);
	\draw [ultra thick, ->]   (3.5,3) -- (7.5,3);
	\draw [ultra thick, -]   (-3.5,0) -- (-3.5,3);
	\draw [ultra thick, <-]   (-7.5,3) -- (-3.5,3);
	
	\node [above right] at (0,4) {$V(x)$};
	\node [right] at (0.75,3) {$V_0$};
	\node [right] at (7,0) {$x$};
	\node [below] at (0,0) {$0$};
	\node [below] at (3.6,0) {$w+b/2$};
	\node [below] at (-3.6,0) {$-(w+b/2)$};
	\node [below] at (-0.8,0) {$-b/2$};
	\node [below] at (0.7,0) {$+b/2$};

\end{tikzpicture}
\caption{A simple double-well potential, consisting of two wells, each of depth $V_0$ and width $w$, separated
by a barrier of height $V_0$ and width $b$. Unlike the potential depicted in Fig.~\ref{Fig1}, the potential of height $V_0$
extends to $x \rightarrow \pm \infty$ beyond the double-well region.}
\label{fig:a1}
\end{figure}
\end{widetext}

This simple double-well potential consists of two wells separated by a barrier of height 
$V_0$ and width $b$ (as in the potential
illustrated in the main body of the text in Fig.~\ref{Fig1}), but with outside constant potential regions that continue to infinity on both sides.
Moreover, these wells are centered around $x=0$ and each have width $w$.
The analytical form is\cite{remark3}
\begin{equation}
V_{\text sq}(x) = \begin{cases} 0 &
\text{if $\left| |x| -({b + w \over 2}) \right| <  {w \over 2}$} \\ 
V_0 & \text{otherwise;}
\end{cases}
\label{square_doublewell_potential} 
\end{equation}
this potential is illustrated in Fig.~A1. Bound state solutions are categorized as either even or odd about the central
barrier. A piecewise-continuous wave function is required over five different regions; then a straightforward matching 
of the wave function and its derivative at the potential discontinuities yields the result
\begin{equation}
 \left( \cos{\frac{qw}{2}} - \frac{q}{\kappa_2}\sin{\frac{qw}{2}}\right) 
\left( \cos{\frac{qw}{2}} + \frac{\kappa_2}{q}\sin{\frac{qw}{2}} \right)
= \pm \tilde{\eta}
\label{eq:nuapp}
\end{equation}
where, as in the body of this paper, $q \equiv \sqrt{2mE/\hbar^2}$ and $\kappa_2 \equiv \sqrt{2m(V_0 - E)/\hbar^2}$. The
positive (negative) sign refers to the even (odd) parity solution. This
result is very similar to Eq.~(\ref{eq:nu}) except here the RHS can take on only two values, with
$\tilde{\eta} \equiv k_0^2 {\rm sin}(qw/2) {\rm cos}(qw/2) \, e^{-\kappa_2 b}$. Now when there is no coupling 
between the wells
(i.e.~$b \rightarrow \infty$ and therefore $\tilde{\eta} \rightarrow 0$), then the vanishing of the first (second) factor on the LHS 
corresponds to determining the energy for the even (odd) bound states in the single well, as was the case
in Eq.~(\ref{eq:nu}). Using the same dimensionless parameters as in the Kronig-Penney case, we define
${z} \equiv q w/2$ and $z_0 \equiv k_0 w/2$, where $k_0 \equiv \sqrt{2mV_0/\hbar^2}$. Then
Eq.~(\ref{eq:nuapp}) becomes
\begin{widetext}
\begin{equation}
\left( \cos{{z}} - \frac{{z}}{\sqrt{z_0^2 - {z}^2}} \sin{{z}} \right) 
\left( \cos{{z}} + \frac{\sqrt{z_0^2 - {z}^2}}{{z}}\sin{{z}} \right) = 
\pm {z_0^2 \over z_0^2 - {z}^2} \sin{{z}} \cos{{z}} \ e^{- \frac{2b}{w} \sqrt{z_0^2 - {{z}}^2}} 
\label{eq:nuzapp}
\end{equation}
\end{widetext}
where we have used $\kappa_2 w/2 = \sqrt{z_0^2 - {z}^2}$. 

As in the Kronig-Penney case, the zeroth order solution is given by $\tilde{z}_1$ (see Eq.~(\ref{eq:cosdelta}) or (\ref{eq:tanz})),
and the solution to Eq.~(\ref{eq:nuzapp}) can be obtained to $1^\text{st}$ order in $\tilde{\eta}$ by writing $z = \tilde{z}_1 + \tilde{\rho}$. With algebra
similar to that which produced Eq.~(\ref{1st_order}), we obtain a result very similar to that equation:
\begin{equation}
\tilde{\rho} = \mp \frac{\tilde{z}_1}{z_0^2} 
\frac{(z_0^2 - \tilde{z}_1^2)}{(1 + \sqrt{z_0^2 - \tilde{z}_1^2})} e^{- \frac{2b}{w} \sqrt{z_0^2 - \tilde{z}_1^2}}.
\label{1st_order_app}
\end{equation}
As expected, the negative (positive) result is precisely {\it half} the value give by Eq.~(\ref{1st_order}) with $k = 0$ ($k = \pi/\ell$),
a result well known for tight-binding models when only two sites (without periodic boundary conditions) are used.

Therefore a toy model with two states only, corresponding to ``particle in left well'' and ``particle in right well'', each with energy
$E_{b} = 4\tilde{z}_1^2 E_0$ (see Eq.~(\ref{eq:tightbindinganal})), that has a tunneling amplitude $\tilde{t}$ for one of these
two degenerate states to tunnel into the other (analogous to the $t$ in Eq.~(\ref{feynman})) then results in two states
with non-degenerate energies, $E_{b} \mp \tilde{t}$. The parameter $\tilde{t}$ is given by the same value as in the
Kronig-Penney model, Eq.~(\ref{e1}), reproduced here for convenience:
\begin{eqnarray}
\tilde{t} = t_1 &=&8 E_0 \tilde{z}_1^2 \frac{1 -\tilde{\delta}^2}{1+z_0 \sqrt{1 - \tilde{\delta}^2}} 
e^{-2 \frac{b}{w} z_0 \sqrt{1 - \tilde{\delta}^2}} \nonumber \\
 &=& 8E_0 z_0 \tilde{\delta}^2 f_1 e^{-x},
\label{eq:t1app}
\end{eqnarray}
with $\tilde{\delta} \equiv \tilde{z}_1/z_0$, $x \equiv 2 \frac{b}{w} \sqrt{z_0^2 - \tilde{z}_1^2}$, and
\begin{equation}
f_1 = {1 -\tilde{\delta}^2 \over \sqrt{1 -  \tilde{\delta}^2} + {1 \over z_0}}.
\label{f1_appa}
\end{equation}

It is clear that this double-well potential (as opposed to the one discussed in Sec.~\ref{sec:doubwell}) more naturally generalizes
to the Kronig-Penney model described in Sec.~\ref{sec:kp}.

Since good agreement with the exact results will be seen to require higher-order corrections (see Sec.~\ref{sec:kp} and Appendix B),
we state the result here as well. 

\begin{widetext}
\phantom{a}
\begin{figure}[h]
\includegraphics[width=4.0cm,angle=-90]{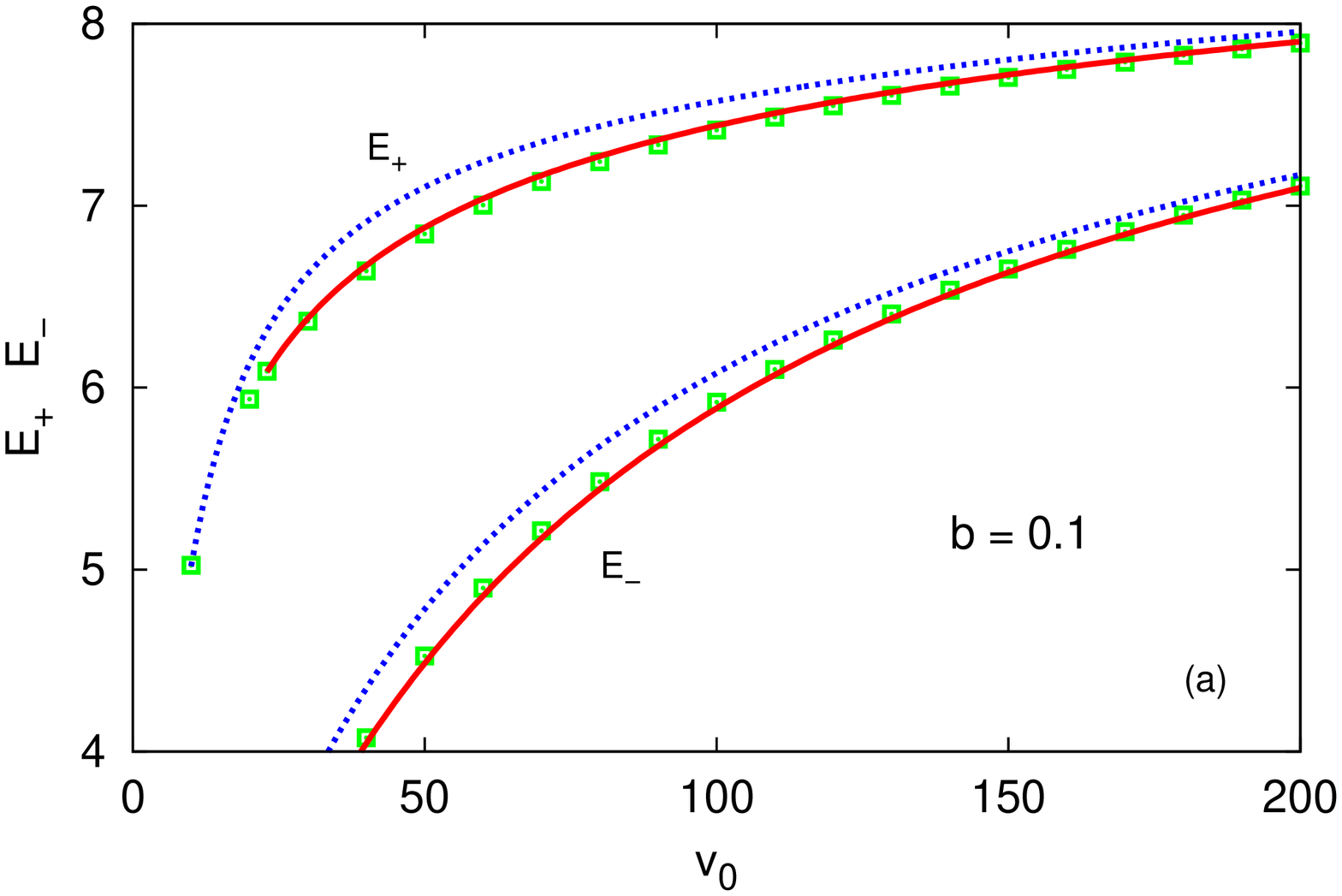}
\includegraphics[width=4.0cm,angle=-90]{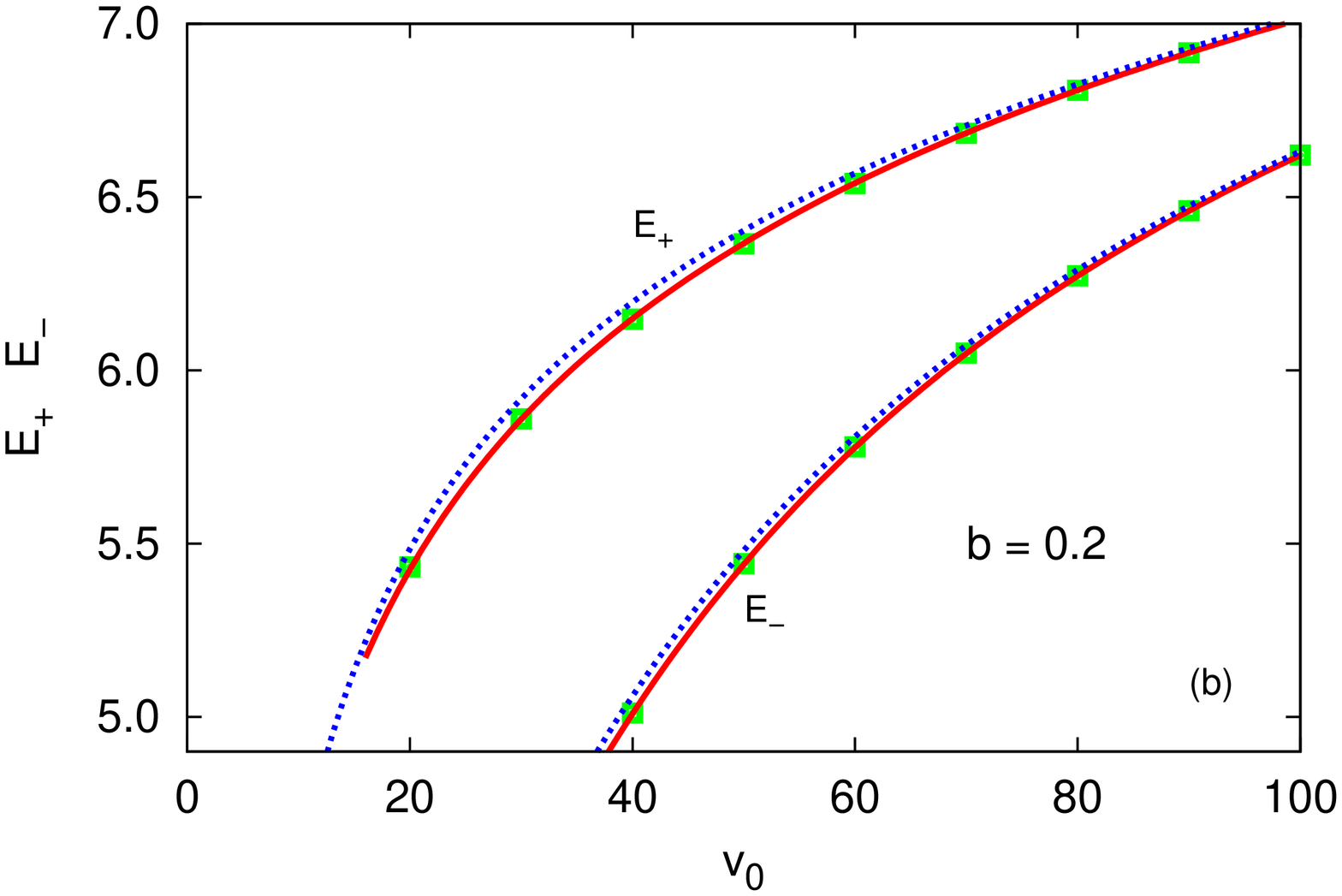}
\includegraphics[width=4.0cm,angle=-90]{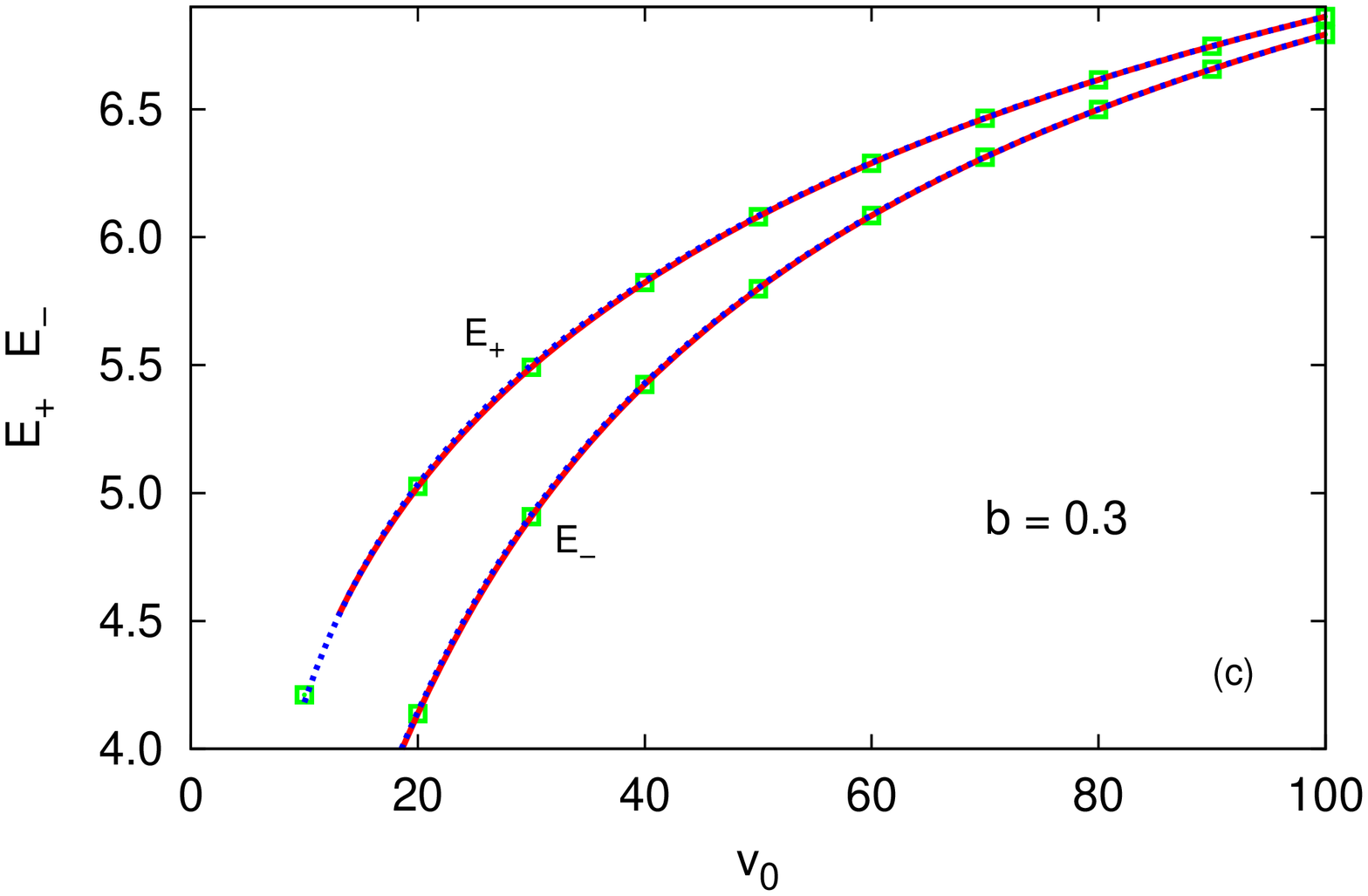}
\caption{Comparison of the first-order result given by $E_{\pm} = E_b \pm \tilde{t}$ (dashed blue curve) and the 
$2^\text{nd}$-order result given by Eq.~(\ref{eb}) (green squares) with the exact result determined numerically from Eq.~(\ref{eq:nuzapp})
(solid red curve) as a function of $v_0 \equiv V_0/E_0$ for (a) $b = 0.1$, (b) $b = 0.2$, and (c) $b = 0.3$, with the well width
adjusted so that $w = 1 - b$ (to conform with the Kronig-Penney case treated in Sec.~\ref{sec:kp} and in Appendix B). The $2^\text{nd}$-order solution is accurate in all three cases, but the $1^\text{st}$-order solution is accurate only when the two wells are
sufficiently far apart [case (c), where essentially all of the curves and points agree].
}
\label{figa2}
\end{figure}
\end{widetext}
With 
\begin{equation}
E_{\pm} = E_b^{(2)} \mp \tilde{t} 
\label{2nd}
\end{equation}
with the superscript `(2)' referring to the fact that $2^\text{nd}$ order corrections in $e^{-x}$ are now included, we find that $\tilde{t}$
is unchanged from the previous result [Eq.~(\ref{eq:t1app})], but the base term, $E_b^{(2)}$, becomes
\begin{equation}
E_b^{(2)} = 4\tilde{z}_1^2\left( 1 - {2f_1^2 \over z_0}e^{-2x}f_2\right),
\label{eb}
\end{equation}
where
\begin{equation}
f_2 = {1 - 2\tilde{\delta}^2 \over \sqrt{1 - \tilde{\delta}^2}} - {1 \over 2z_0} - {2b \over w} {\tilde{\delta}^2 \over \sqrt{1 - \tilde{\delta}^2}}
+ {1 \over z_0^4} {\tilde{\delta}^2 \over 1 - \tilde{\delta}^2} {1 - z_0^2 \tilde{\delta}^2/2 \over \sqrt{1 - \tilde{\delta}^2} + {1 \over z_0}}.
\label{f2}
\end{equation}
Note that the first term in $f_2$ is of order unity but subsequent terms are of lower order in $1/z_0$. Fig.~\ref{figa2} shows the two
split (lowest) energies for three examples ($b = 0.1$, $b = 0.2$ and $b = 0.3$, with $w \equiv 1 - b$ to conform with the Kronig-Penney
parameters). Note that the energies with $2^\text{nd}$ order corrections are in good agreement for essentially all values of $V_0$ for
all three cases, whereas the energies calculated with $1^\text{st}$-order corrections only agree with the exact solutions only in case (c)
or for larger values of $v_0$ than shown here. All of these higher order corrections occur due to the inherent nonlinear nature of
Eq.~(\ref{eq:nuzapp}), i.e.~they do {\it not} occur because of next-nearest-neighbor tunneling (since there are only two wells
and hence no next-nearest neighbors). 

\section{Derivation of Eq.~\ref{1st_order}.}

We begin with Eq.~(\ref{eq:nuz}), but with the higher order correction, $\eta_2$ omitted. Then,
\begin{widetext}
\begin{equation}
\left( \cos{z} - \frac{z}{\sqrt{z_0^2 - z^2}} \sin{z} \right) 
\left( \cos{z} + \frac{\sqrt{z_0^2 - z^2}}{z}\sin{z} \right) \approx 2 e^{-2 \frac{b}{w} \sqrt{z_0^2 - \tilde{z}_1^2}} \cos{k \ell}.
\label{eq:nuz_appb}
\end{equation}
\end{widetext}
Since the right-hand-side (RHS) is exponentially small, and we will pursue the dispersion for the lowest (even) bound state,
then the solution is given by 
\begin{equation}
z =  \tilde{z}_1 \left[ 1 + \tilde{\rho}(k)\right],
\label{z_app1}
\end{equation}
where $\tilde{\rho}(k)$ is a small relative correction to the solution for a single well, denoted by $\tilde{z}_1$, and determined
by the first factor on the left-hand-side (LHS) of Eq.~(\ref{eq:nuz_appb}) being zero. 
That is, Eq.~(\ref{eq:cosdelta}) determines  $\tilde{z}_1$.

Inserting the solution given by Eq.~(\ref{z_app1}) and expanding the first factor of of Eq.~(\ref{eq:nuz_appb})
to first order in $\tilde{\rho}(k)$, gives for this first factor
\begin{widetext}
\begin{equation}
\left( \cos{z} - \frac{z}{\sqrt{z_0^2 - z^2}} \sin{z} \right)  \approx -\tilde{\rho}(k) \left( \sqrt{z_0^2 - \tilde{z}_1^2} + 1 \right)
{z_0^2 \cos{\tilde{z}_1} \over z_0^2 - \tilde{z}_1^2}.
\label{expansiona}
\end{equation}
\end{widetext}

Since Eq.~(\ref{expansiona}) is already first order in $\tilde{\rho}(k)$ (as it must be), then the second factor on the LHS 
of Eq.~(\ref{eq:nuz_appb}) is required only to zeroth order; we readily obtain
\begin{equation}
\left( \cos{z} + \frac{\sqrt{z_0^2 - z^2}}{z}\sin{z} \right) \approx {z_0^2 \over \tilde{z}_1^2} \cos{\tilde{z}_1}.
\label{expansionb}
\end{equation}
Taking the product of Eq.~(\ref{expansiona}) and Eq.~(\ref{expansionb}) gives
\begin{equation}
{\rm LHS} \approx -\tilde{\rho}(k) \left( \sqrt{z_0^2 - \tilde{z}_1^2} + 1 \right)
{z_0^2 \over z_0^2 - \tilde{z}_1^2}.
\label{product}
\end{equation}
where we have used $\cos^2{\tilde{z}_1} = {\tilde{z}_1^2/z_0^2}$. Equating this to the RHS of Eq.~(\ref{eq:nuz_appb})
then gives Eq.~(\ref{1st_order}).

\section{2nd order in the exponential}

For convenience, we repeat Eq.~(\ref{eq:nuz}) and the two auxiliary equations:
\begin{widetext}
\begin{equation}
\left( \cos{z} - \frac{z}{\sqrt{z_0^2 - z^2}} \sin{z} \right) 
\left( \cos{z} + \frac{\sqrt{z_0^2 - z^2}}{z}\sin{z} \right) = \eta_1(k) + \eta_2,
\label{eq:nuza}
\end{equation}
\begin{equation}
\eta_1(k) = 2 e^{-2 \frac{b}{w} \sqrt{z_0^2 - \tilde{z}_1^2}} \cos{k \ell}
\label{eq:nuz1a}
\end{equation}
and
\begin{equation}
\eta_2 = e^{-4 \frac{b}{w} \sqrt{z_0^2 - \tilde{z}_1^2}} \left( \cos{z} - \frac{\sqrt{z_0^2 - z^2}}{z} \sin{z} \right) 
\left( \cos{z} + \frac{z}{\sqrt{z_0^2 - z^2}}\sin{z} \right).
\label{eq:nuz2a}
\end{equation}
As stated in the text these are exact. In the tight-binding limit the wells are infinitely separated, so $b \rightarrow \infty$,
and therefore the RHS of Eq.~(\ref{eq:nuza}) is expected to be small. We therefore proceed as in the text, and write
the solution as $z \approx \tilde{z}_1\left(1 + \tilde{\rho}(k)\right)$, and expand to $2^{\rm nd}$ order in $\tilde{\rho}(k)$ as
well as in $e^{-x}$, where $x \equiv 2 \frac{b}{w} \sqrt{z_0^2 - \tilde{z}_1^2}$, since $\eta_2/\eta_1(k) \approx O(e^{-x})$.
A straightforward calculation gives
\begin{equation}
-\tilde{\rho}(k){ z_0 \over f_1} \left\{1 - \tilde{\rho}(k) g_1\right\} = 
2 e^{-x} \cos{k \ell} + 2z_0 {2 b \over w} e^{-x} \tilde{\rho}(k) {\tilde{\delta}^2 \over \sqrt{1 - \tilde{\delta}^2}} \cos{k \ell} + 2 e^{-2x}
\left(1 - 2 \tilde{\delta}^2 \right),
\label{2nd_rho_1}
\end{equation}
\end{widetext}
where 
\begin{eqnarray}
f_1 &=& {1 -\tilde{\delta}^2 \over \sqrt{1 -  \tilde{\delta}^2} + {1 \over z_0}} \nonumber \\
g_1 &=& 1 - {\tilde{\delta}^2 \over 1 - \tilde{\delta}^2} { \sqrt{1 - \tilde{\delta}^2}
 + {3 \over 2z_0} \over \sqrt{1 - \tilde{\delta}^2}  + {1 \over z_0}},
\label{fgdefns}
\end{eqnarray}
and $\tilde{\delta} \equiv \tilde{z}_1/z_0$. Eq.~(\ref{2nd_rho_1}) is a quadratic in $\tilde{\rho}(k)$; since this is a small quantity we can
simply iterate to obtain $\tilde{\rho}(k)$ explicitly:
\begin{eqnarray}
\tilde{\rho}(k) &=& -{2 \over z_0} f_1e^{-x} \cos{k \ell} \nonumber \\
&& - {2f_1 \over z_0} e^{-2x}\left( 1 - 2 \tilde{\delta}^2  - {f_1 \over z_0} \left( g_1 + z_0 {2 b \over w} 
{\tilde{\delta}^2 \over \sqrt{1 - \tilde{\delta}^2}} \right) \right) \nonumber \\
&& + {2f_1^2 \over z_0^2} e^{-2x} \left( g_1 + z_0 {2 b \over w} 
{\tilde{\delta}^2 \over \sqrt{1 - \tilde{\delta}^2}}  \right) \cos{2k \ell}.
\label{2nd_rho_2}
\end{eqnarray}
Finally, we use
\begin{equation}
E(k) = 4E_0 \tilde{z}_1^2(1 + \tilde{\rho})^2
\label{enera}
\end{equation}
from which we obtain
\begin{equation}
E(k) = E_c -2t_1 \cos{k \ell} - 2t_2 \cos{2k \ell},
\label{ener_2ndorder}
\end{equation}
where the parameters defined by Eq.~(\ref{ener_2ndorder}) are given by
\begin{equation}
t_1 = 8E_0 z_0 \tilde{\delta}^2 f_1 e^{-x},
\label{e1}
\end{equation}
\begin{equation}
t_2 = -8E_0 \tilde{\delta}^2 f_1^2\left(g_1 + {1 \over 2} + z_0 {2 b \over w} {\tilde{\delta}^2 \over \sqrt{1 - \tilde{\delta}^2}}\right)e^{-2x}
\label{e2}
\end{equation}
and
\begin{equation}
E_c = E_0\left\{ 4\tilde{z}_1^2 - 16\tilde{z}_1 \tilde{\delta} f_1 e^{-2x}  \left( 1 - 2 \tilde{\delta}^2 \right) \right\} - 2t_2.
\label{e0}
\end{equation}
Note that the expansion is governed by the exponential suppression contained in the $e^{-x}$ and $e^{-2x}$ factors. However,
we expect $z_0 > 1$ for tight-binding, whereas Fig.~\ref{Fig4} makes it clear that $\tilde{z}_1$ is of order unity or lower; more precisely, 
$\tilde{z}_1 < {\rm min}(z_0, \pi/2)$, and so $\tilde{\delta} <1$ as well. We have written the above 
expressions to make these expansion parameters clear; thus, even within
the expression for $t_2$, for example, various terms will contribute significantly less than others. Both $f_1$ and $g_1$ are
of order unity.

The result from Eq.~(\ref{ener_2ndorder}) is plotted in Fig.~\ref{Fig5} (as indicated by the square symbols), 
and gives a very accurate result for the parameters used in that figure.
Note that $t_1$ is given by an expression identical to the one implied in Eq.~(\ref{eq:tightbindinganal})---by going to
$2^\text{nd}$ order in $e^{-x}$ this has not changed, and in fact is identical to the expression derived for the double-well in
Appendix A, Eq.~(\ref{eq:t1app}). The need to to go to $2^\text{nd}$ order and therefore generate a term with
$\cos{(2k\ell)}$ wave vector dependence is sometimes interpreted to mean that a significant tunneling amplitude exists between
second-nearest-neighbor atoms. The fact that this is required even for the case of the double-well studied in Appendix A
(where there is no second-nearest neighbor!) indicates that this interpretation is incorrect.


\begin{thebibliography}{99}
\bibitem{remark1} We use the trigonometric identities $\cos qw = \cos^2{\left({qw}/{2}\right)} - \sin^2{\left({qw}/{2}\right)}$ and 
$\sin qw = 2 \sin{\left({qw}/{2}\right)}\cos{\left({qw}/{2}\right)}$.
\bibitem{remark3} Note that the form is identical to that in Ref.~[\onlinecite{dauphinee15}] but our definitions of certain
parameters differ from those used there.
\end{thebibliography}
\end{document}